\documentclass[12pt]{article}
\usepackage{graphicx,epsfig}
\usepackage{color}
\usepackage{cite}
\title{The influence of nonlocal interactions on valence transitions and formation  
of excitonic bound states in the generalized Falicov-Kimball model}
\author{Pavol Farka\v sovsk\'y\\
Institute  of  Experimental  Physics,  Slovak   Academy   of
Sciences\\
Watsonova 47, 043 53 Ko\v {s}ice, Slovakia}
\date{}
\begin{document}
\baselineskip=24pt
\maketitle

\begin{abstract}
We use the density-matrix-renormalization-group (DMRG) method to study 
the combined effects of nonlocal interactions on valence transitions 
and the formation of excitonic bound states in the generalized Falicov-Kimball 
model. In particular, we consider the nearest-neighbour Coulomb interaction 
$U_{nn}$ between two $d$, two $f$, $d$ and $f$ electrons as well as the 
so-called correlated hopping term $U_{ch}$ and examine their effects on 
the density of conduction $n_d$ (valence $n_f$) electrons and the excitonic 
momentum distribution $N(q)$. It is shown that $U_{nn}$ and $U_{ch}$ exhibit
very strong and fully different effects on valence transitions and the
formation (condensation) of excitonic bound states. While the nonlocal 
interaction $U_{nn}$ suppresses the formation of zero momentum condensate 
($N(q$=$0)$) and stabilizes the intermediate valence phases with 
$n_d \sim 0.5, n_f \sim 0.5$, the correlated hopping term $U_{ch}$ 
significantly enhances the number of excitons in the zero-momentum 
condensate and suppresses the stability region of intermediate valence 
phases. The physically most interesting results are observed if both 
$U_{nn}$ and $U_{ch}$ are nonzero, when the combined effects of $U_{nn}$ 
and $U_{ch}$ are able to generate discontinuous changes in $n_f$, 
$N(q$=$0)$ and some other ground-state quantities. 

\end{abstract}

\newpage
\section{Introduction}
Since its introduction in 1969 the Falicov-Kimball model
has become an important standard model for a description of   
correlated fermions on the lattice~\cite{Falicov}. It has been used
in the literature to study a great variety of many-body effects    
in metals, of which charge-density waves, metal-insulator transitions 
and mixed-valence phenomena are the most common examples~\cite{Cho}.
In the past years the Falicov-Kimball model was extensively studied 
in connection with the exciting idea of electronic 
ferroelectricity~\cite{P1,P2,Cz,F1,F2,Zl,B1,B2,F3,Schneider},
which is directly related with the formation and condensation of 
excitonic bound states of conduction ($d$) and valence ($f$) 
electrons~\cite{Z1,Phan,Seki,Z2,Kaneko1,Kaneko2,Ejima}. In its 
original form, the Falicov-Kimball model describes a two-band system of
localized $f$ and itinerant $d$ electrons with short-ranged
$f$-$d$ Coulomb interaction $U$:
\begin{equation}
H_0=\sum_{ij}t_{ij}d^+_id_j+U\sum_if^+_if_id^+_id_i+E_f\sum_if^+_if_i,
\end{equation}
where $f^+_i$, $f_i$ are the creation and annihilation
operators  for an electron in  the localized state at 
lattice site $i$ with binding energy $E_f$ and $d^+_i$,
$d_i$ are the creation and annihilation operators
of the itinerant spinless electrons in the $d$-band
Wannier state at site $i$. 

The first term of (1) is the kinetic energy
corresponding to quantum-mechanical hopping of the itinerant $d$ electrons
between sites $i$ and $j$. These intersite hopping
transitions are described by the matrix  elements $t_{ij}$,
which are $-t_d$ if $i$ and $j$ are the nearest neighbors and
zero otherwise (in the following all parameters are measured 
in units of $t_d$). The second term represents the on-site   
Coulomb interaction between the $d$-band electrons with density
$n_d=\frac{1}{L}\sum_id^+_id_i$ and the localized
$f$ electrons with density $n_f=\frac{1}{L}\sum_if^+_if_i$,
where $L$ is the number of lattice sites. The third  term stands
for the localized $f$ electrons whose sharp energy level is $E_f$.

Since the local $f$-electron  number $f^+_if_i$ is strictly 
conserved quantity, the $d$-$f$ electron coherence cannot be
established in this model. This shortcoming can be overcome
by including an explicit local hybridization     
$H_V=V\sum_id^+_if_i+f^+_id_i$ between the $d$   
and $f$ orbitals. This model has been extensively
studied in our previous work~\cite{Fark_epl}.
The numerical analysis of the excitonic momentum distribution 
$N(q)=\langle b^+_qb_q\rangle$ (with $b^+_q=(1/\sqrt{L})\sum_k
d^+_{k+q}f_k$)
showed that this quantity diverges for $q=0$, signalizing a
Bose-Einstein condensation of preformed excitons. The stability   
of the zero-momentum ($q=0$) condensate against the $f$-electron  
hopping has been studied in our very recent paper~\cite{Fark_prb}.
It was found that the negative values of the $f$-electron hopping 
integrals $t_f$ support the formation of zero-momentum condensate,
while the positive values of $t_f$ have the fully opposite effect.
Moreover, we have found that the fully opposite effects on the
formation of condensate exhibit also the local and nonlocal 
hybridization with an inversion symmetry.  The first one strongly   
supports  the formation of condensate, while the second one destroys
it completely. In systems with equal parity orbitals, the local and 
nonlocal hybridization can coexist and for this case we have 
found~\cite{Fark_ssc} that
the combined effect of both hybridizations strongly support 
the formation of zero-momentum condensate the existence of which
can yield the reasonable explanation for the pressure-induced 
resistivity anomaly observed experimentally in $TmSe_{0.45}Te_{0.55}$ 
compound~\cite{Wachter}.

These results show, that the Falicov-Kimball model has a great potential 
to describe some of anomalous features of real complex materials
like rare-earth compounds. On the other hand it should be noted
that the original version of the model, as well as its extensions
discussed above, represent too crude approximation of real rare-earth
compounds, since neglect all nonlocal Coulomb interactions, that
can change this picture. For the correct description of these materials
one should take into account at least the following nonlocal
Coulomb interaction terms:
\begin{equation}
H_{non}=
U_{dd}\sum_{<ij>}n^d_in^d_j+
U_{df}\sum_{<ij>}n^d_in^f_j+
U_{ff}\sum_{<ij>}n^f_in^f_j+
U_{ch}\sum_{<ij>}d^+_id_j(n^f_i+n^f_j),
\end{equation}
which represent  the nearest-neighbour Coulomb interaction 
between two $d$ electrons (the first term), between one $d$ and one $f$ 
electron (the second term), between two $f$ electrons (the third term)
and the so-called correlated hopping (the last term).

There is a number of papers, were the influence of individual
interaction terms from (2) on the ground state properties
of the Falicov-Kimball model has been studied (see Ref.~\cite{Fark_aps} 
and references therein), however, there are only few, where the combined 
effects of two or three terms were considered~\cite{Fark_app,Lemanski}. 
From this point of view the model Hamiltonian
studied in this paper $H=H_0+H_V+H_{non}$, represents one of the most
complex extensions of the Falicov-Kimball model used for a description
of ground state properties of strongly correlated systems.   
Here we concern our attention exclusively on a description of two 
main problems, and namely, the process of formation and condensation 
of excitonic bound states and the problem of valence transitions 
in the generalized Falicov-Kimball model.

\section{Results and discussion}

To examine effects of the above mentioned factors on the ground states
properties of the generalized Falicov-Kimball model we have performed 
exhaustive DMRG studies of the model Hamiltonian $H=H_0+H_V+H_{non}$ 
for a wide range of the model parameters at the half-filled band
case $n=n_d+n_f=1$. To simplify numerical calculations we adopt here
the following model $U_{dd}=U_{ff}=U_{df}=U_{nn}$, that allows us
to reduce the number of model parameters and at the same time to keep 
all nonlocal interactions terms nonzero. In all our DMRG calculations
we typically keep up  to 500 states per block, although in the numerically 
more difficult cases,where the DMRG results converge slower, we keep up 
to 1000 states. Truncation errors~\cite{White}, given by the 
sum of the density matrix eigenvalues of the discarded states, vary 
from $10^{-6}$ in the worse cases to zero in the best cases. As mentioned 
above the special attention in the current paper is devoted to the
understanding of the process of formation and condensation of excitonic 
bound states. To describe this process in more detail (and at the same 
time to specify the role of different factors) we have calculated numerically 
the density of zero momentum excitons $n_0=\frac{1}{L}N(q=0)$, 
the total exciton density $n_T=\frac{1}{L}\sum_qN(q)$, 
the total $d$-electron density $n_d$ and the total density of 
unbond $d$ electrons $n^{un}_d=n_d-n_T$ as functions of the interaction 
parameters $U_{nn}$ and $U_{ch}$  for zero as well as finite values of $E_f$. 
Let us start our discussion with numerical results obtained for 
the density of zero momentum excitons $n_0$ as function of $U_{nn}$ 
and $U_{ch}$ at $E_f=0$. The numerical results for $n_0$ obtained from the 
DMRG calculations  are summarized in Fig.~1 for $U=1, L=100$ 
and several different values of the local hybridization $V$.
\begin{figure}[h!]
\begin{center}
\includegraphics[width=7.0cm]{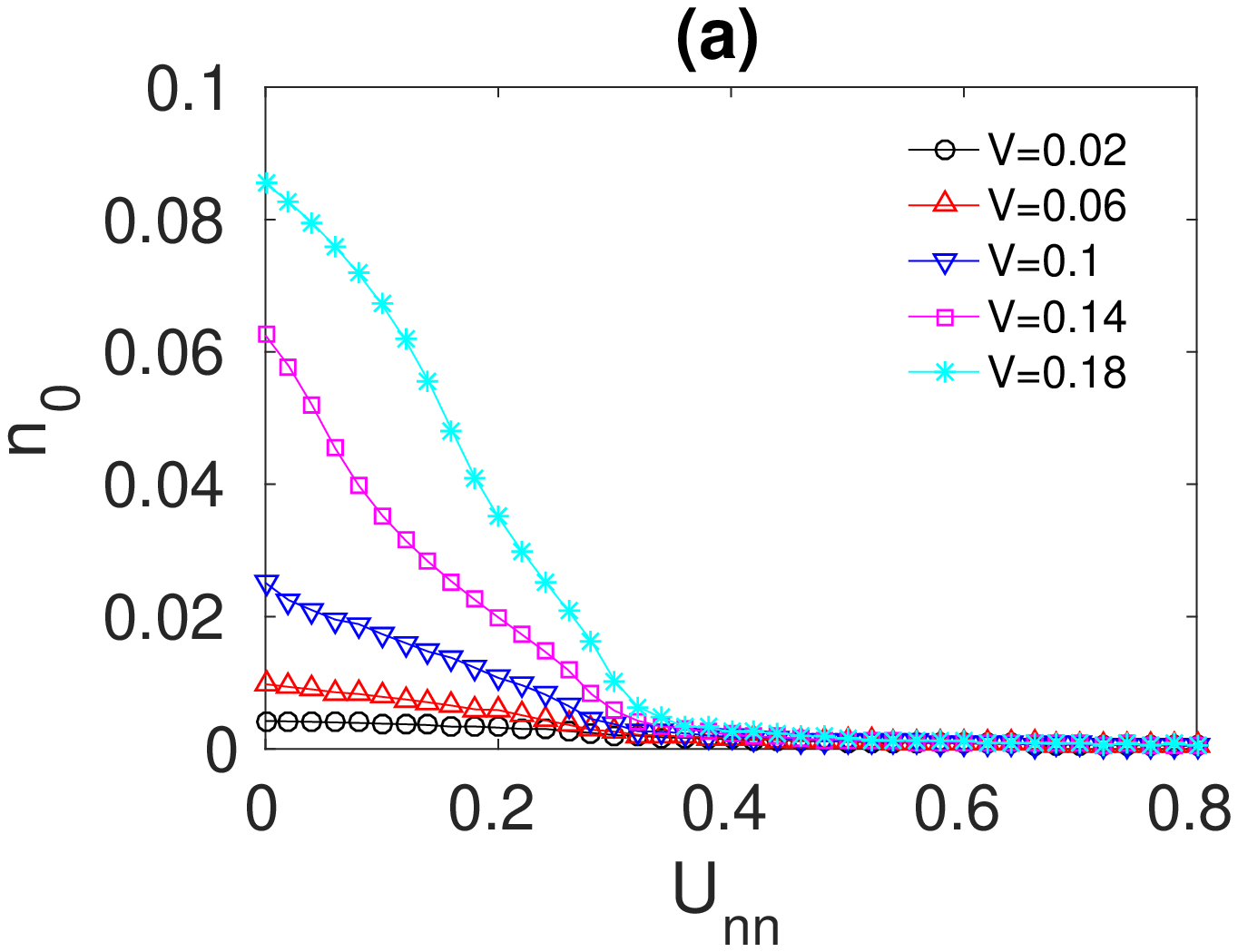}
\includegraphics[width=7.0cm]{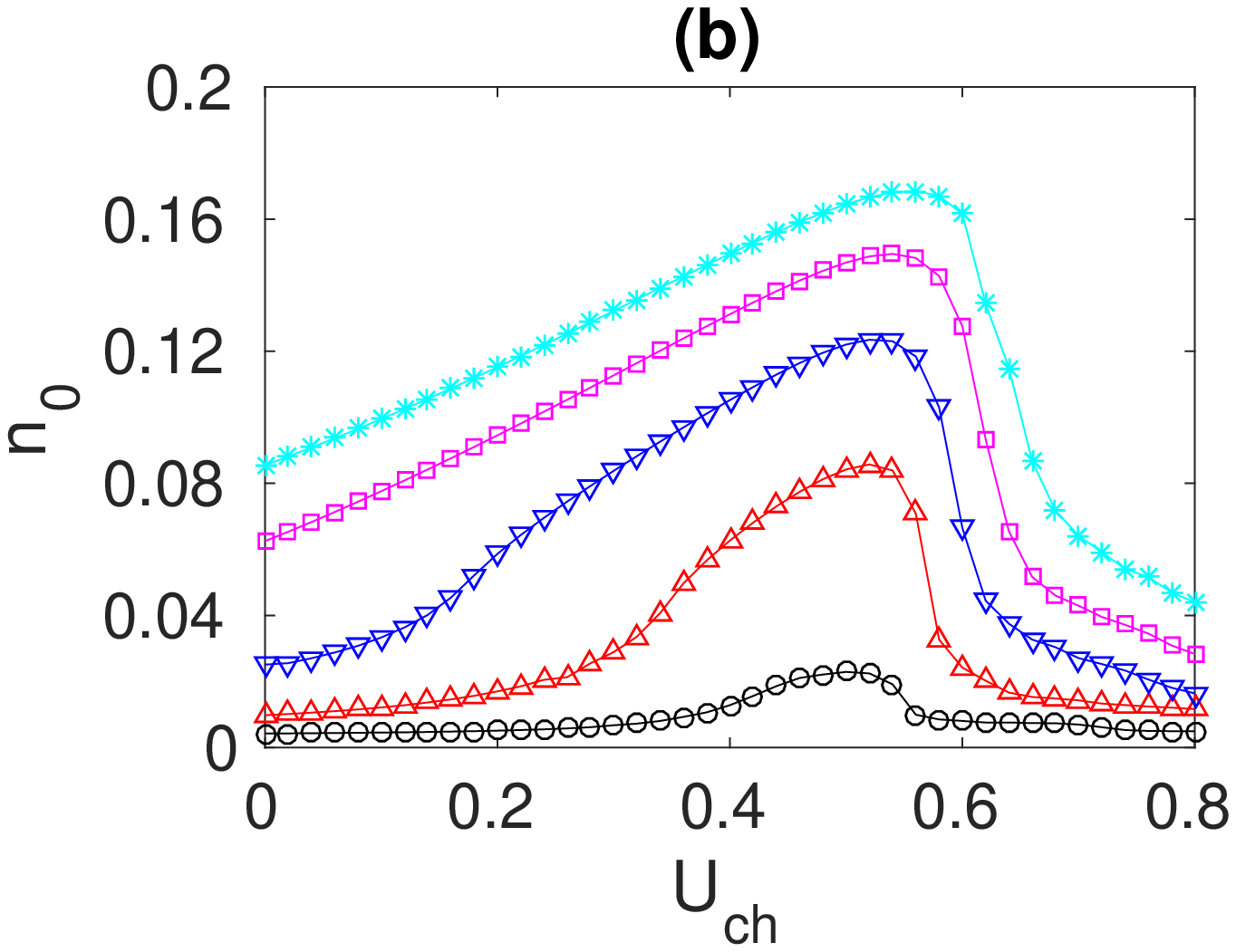}
\includegraphics[width=7.0cm]{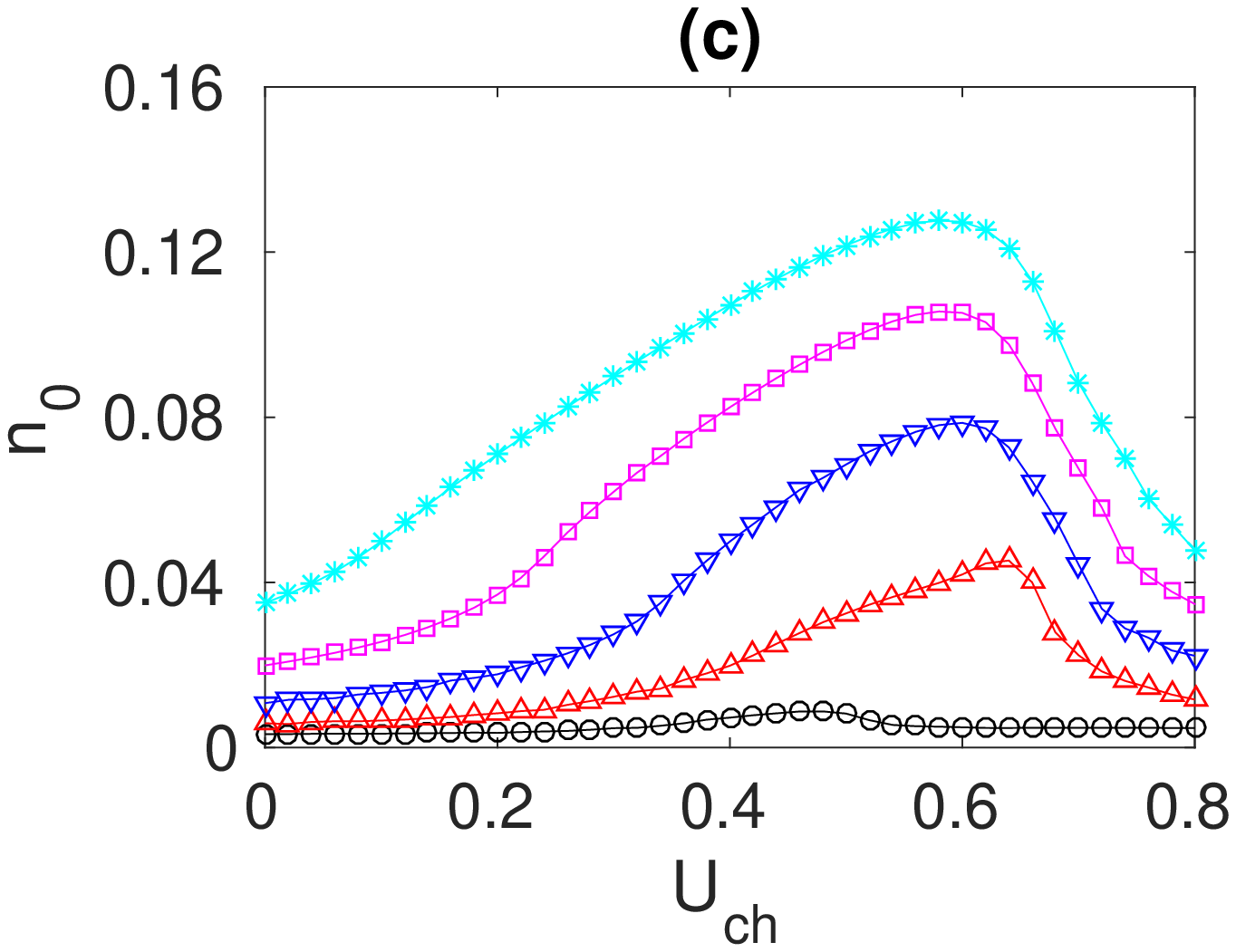}
\includegraphics[width=7.0cm]{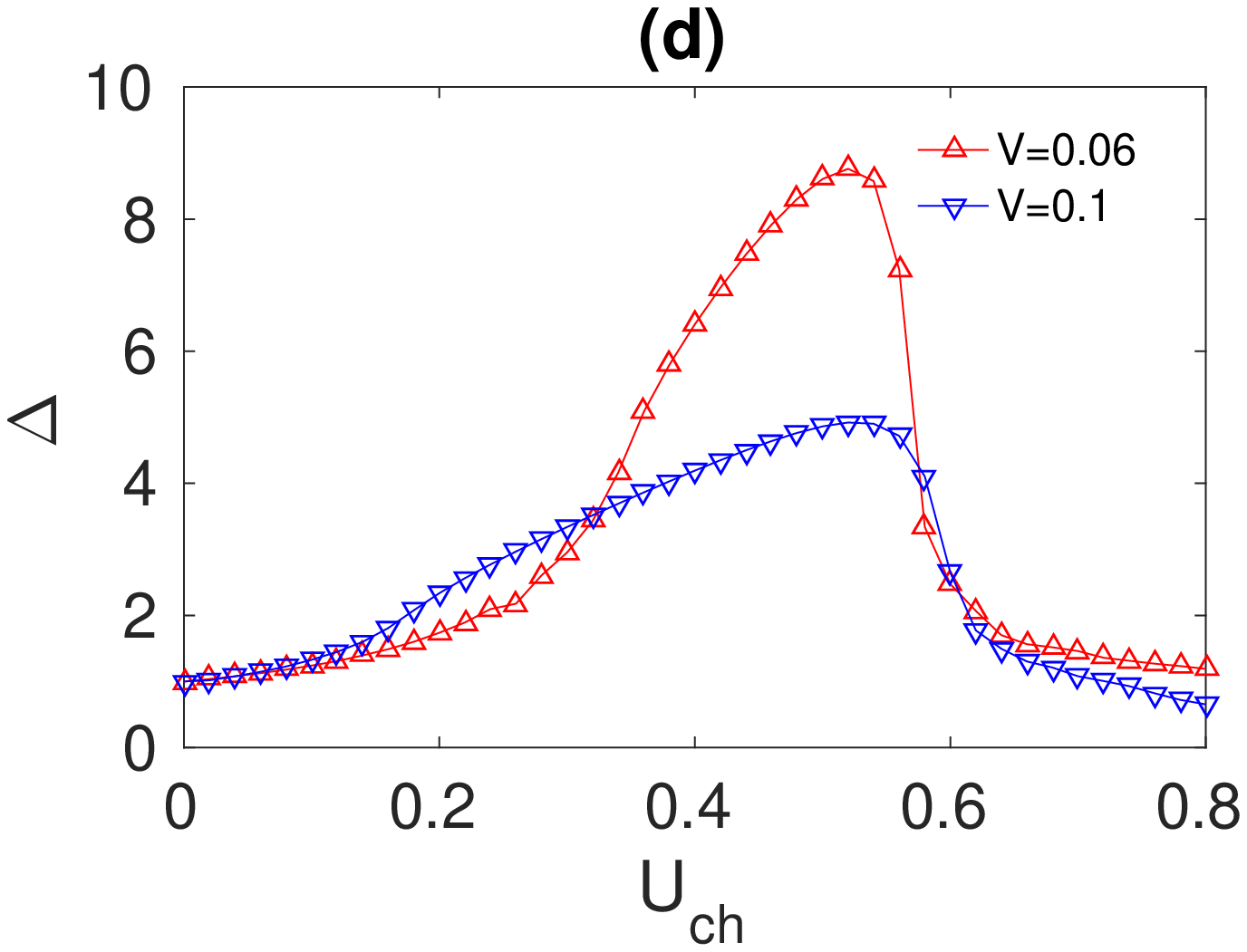}
\end{center}
\caption{\small The density of zero-momentum excitons $n_0$ as a function 
of $U_{nn}$ and $U_{ch}$ calculated for five different values of $V$
($V=0.02, 0.06, 0.14,0.18$) at $U=1, L=100$ and $n_f+n_d=1$. (a) $U_{ch}=0$;
(b) $U_{nn}=0$; (c) $U_{nn}=0.2$; (d) the enhancement 
$\Delta=n_0(U_{ch})/n_0(U_{ch}$=$0)$ as a function of $U_{ch}$ for two different 
values of $V$ at $U_{nn}=0$.}
\label{fig1}
\end{figure}
 These results show that the nonlocal
interactions $U_{nn}$ and $U_{ch}$ exhibit fully different effects
on the condensation of preformed excitons to the zero-momentum
state. The nonlocal interaction $U_{nn}$ strongly suppresses the
condensation of preformed excitons to the ground state and already
relatively small values of $U_{nn}$ ($U_{nn}\sim 0.4$) practically
fully destroy the zero-momentum condensate. On the other hand, the
correlated hopping represented by the $U_{ch}$ interaction term exhibits 
fully opposite effect on $n_0$. In this case the number of zero momentum
excitons is strongly enhanced with increasing $U_{ch}$ and reaches its maximum
at $U^{max}_{ch} \sim 0.6$. The  Fig.~1d shows that the enhancement 
$\Delta=n_0(U_{ch})/n_0(U_{ch}$=$0)$ of the zero momentum excitons due to the 
$U_{ch}$ term is indeed considerable and at the point of the global maximum 
it reaches values $\Delta \sim 5$, for $V=0.6$ and  $\Delta\sim 10$,
for $V=0.1$, pointing on the crucial role of the correlated hopping
in the process of formation and condensation of excitonic bound states. 
Regarding the influence of the local hybridization on the formation
of zero-momentum condensate we have found that in all examined cases 
the nonlocal hybridization does not change qualitatively the behavior 
of $n_0$ and therefore all calculations in the rest of the paper 
are done for the representative value of $V=0.1$.

Till now we have presented results exclusively for $E_f=0$.
Let as now discuss briefly the effect of change of the $f$-level
position. This study is interesting also from this point of 
view that taking into account the parametrization between the
external pressure $p$ and the position of the $f$ level ($E_f \sim p $),
one can deduce from the $E_f$ dependences of the ground state 
characteristics also their $p$ dependences, at least qualitatively~\cite{Gon}.
The resultant $E_f$ dependences of the density of zero momentum 
excitons $n_0$, the total exciton density $n_T$, the total 
$d$-electron density $n_d$ and the total density of unbond 
$d$ electrons $n^{un}_d$ obtained by DMRG method are shown 
in Figs.~2-4 for several values of $U_{nn}$ and $U_{ch}$ at 
$U=1$ and $L=100$. 
\begin{figure}[h!]
\begin{center}
\includegraphics[width=7.0cm]{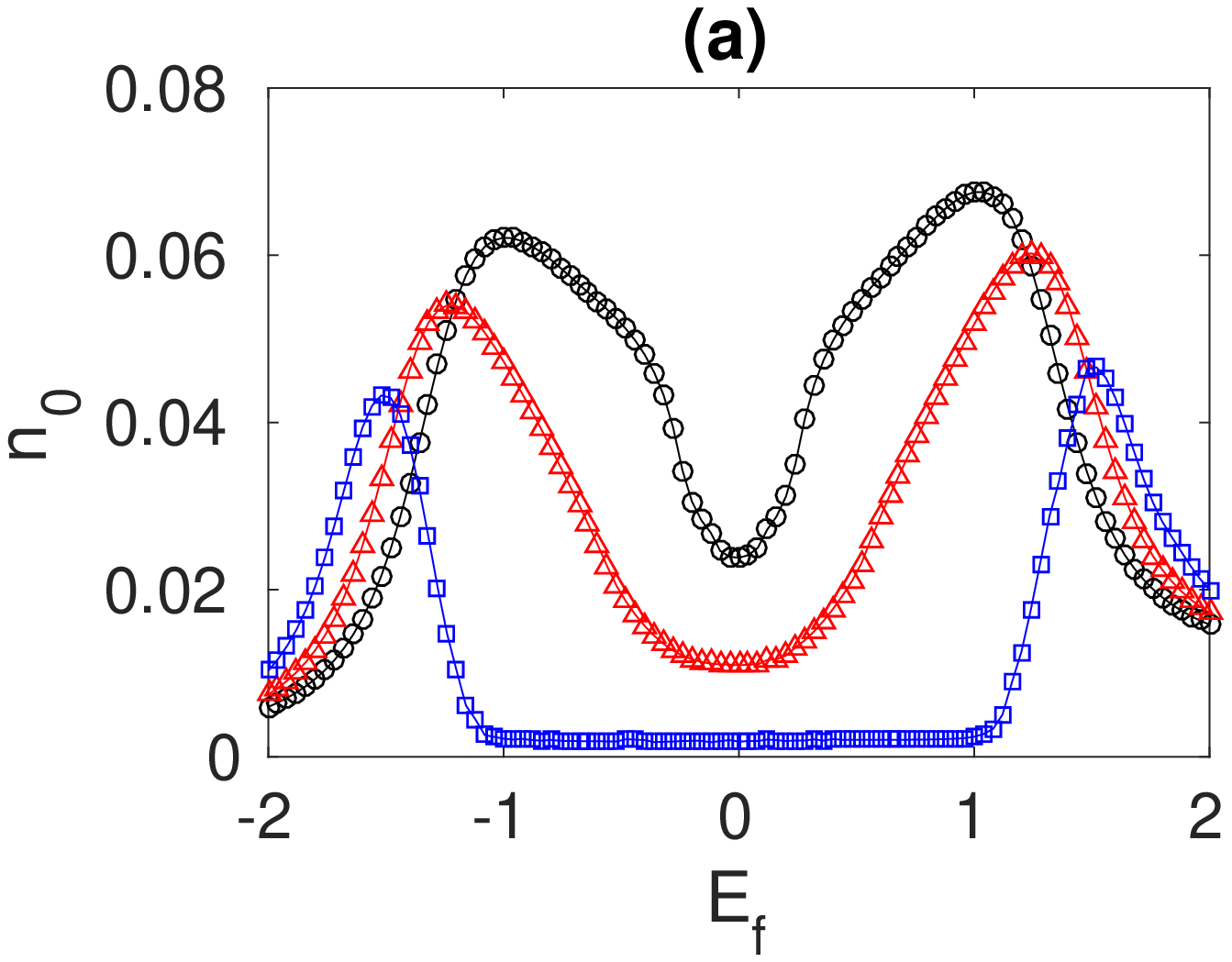}
\includegraphics[width=7.0cm]{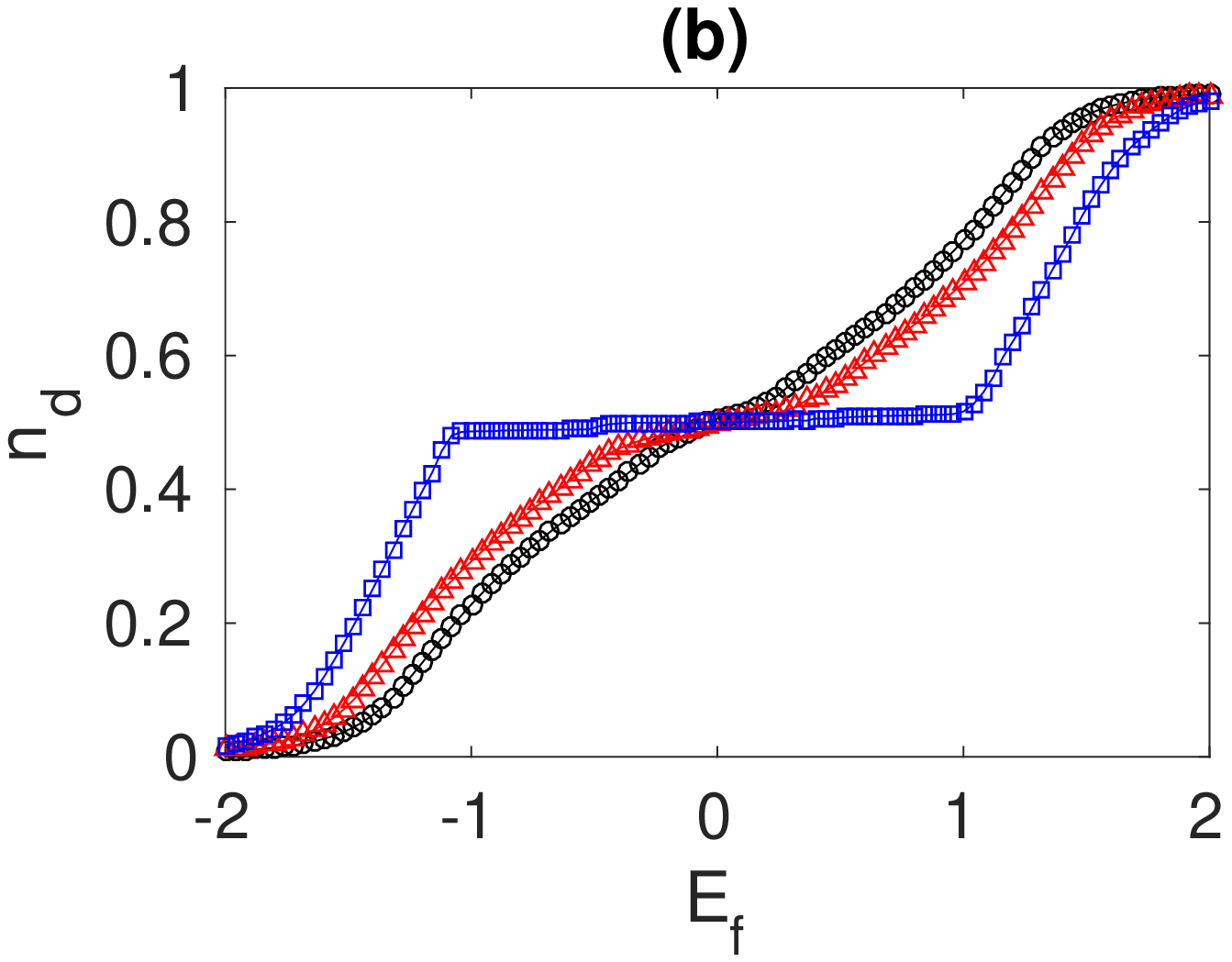}
\includegraphics[width=7.0cm]{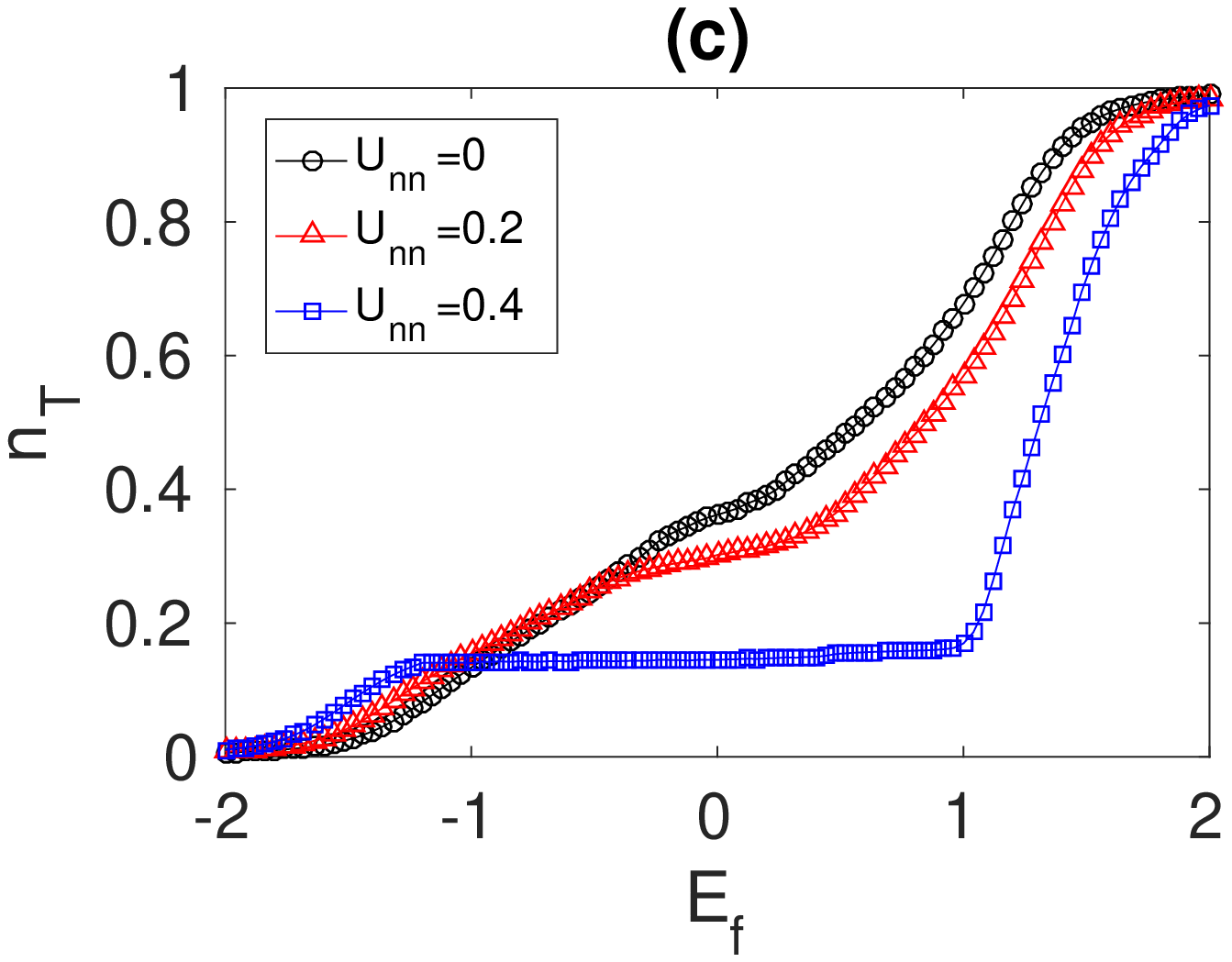}
\includegraphics[width=7.0cm]{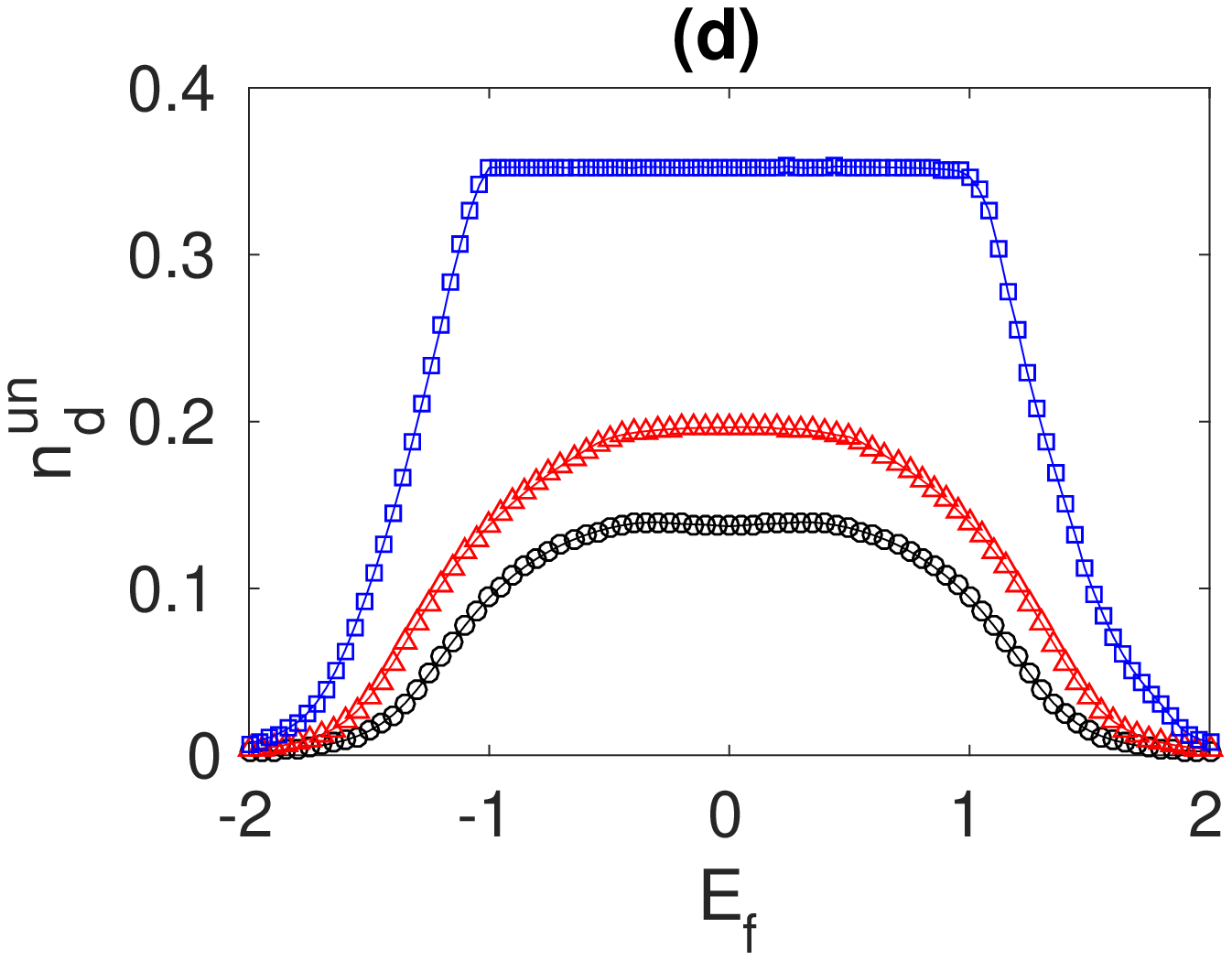}
\end{center}
\caption{\small $n_0, n_d, n_T$ and $n^{un}_d=n_d-n_T$ as functions of $E_f$
calculated for three different values of $U_{nn}$ ($U_{nn}=0, 0.2, 0.4$)
at $U_{ch}=0, U=1, V=0.1, L=100$ and $n_f+n_d=1$.}
\label{fig2}
\end{figure}
Let us first discuss the effect of the $U_{nn}$ term (see Fig.~2a). 
One can see that the density of zero momentum excitons is nonzero over
the whole interval of $E_f$ values and is considerably enhanced 
in some regions,  due to the significant enhancement of the 
$d$ electron population in the $d$ band (see panel b).
Generally the increasing nonlocal interaction $U_{nn}$ destroys the 
zero-momentum condensate and this effect is most pronounced near 
the half-filled band point $n_d=n_f=1/2$, obviously due to the formation
of the charge-density-wave phase that is the ground state of the model
at this point for $U_{nn}$ and $U_{ch}$ equals zero. Another general trend 
recognized from our numerical calculations concerns the valence 
transition behaviour of the model, and particularly, the stabilization 
of intermediate valence plateau at $n_d=n_f=1/2$ with increasing $U_{nn}$.
This effect is very strong and already very small values of the nonlocal
interaction $U_{nn}$ ($U_{nn}=0.4$) are able to stabilized the intermediate
valence  phase with $n_f \sim 1/2$ for a wide range of $E_f$ values.
The total exciton density $n_T$ exhibits qualitatively the same behaviour as
the total $d$-electron density $n_d$, however their 
difference $n_d-n_T$, which represents the total density of unbond 
$d$ electrons, exhibits strong dependence on both $U_{nn}$ as well as $E_f$. 
With increasing $U_{nn}$, the total density of unbond $d$ electrons 
$n^{un}_d$ is generally enhanced, while with increasing $E_f$, the total 
density of unbond $d$ electrons increases for $E_f<0$ and decreases
for $E_f>0$, with obvious tendency to form the plateau around 
$E_f=0$ for $U_{nn}$ sufficiently large (small deviations from this
behaviour are observed only near the point $E_f=0$ for $U_{nn}=0$).

In comparison to $U_{nn}$ effects, the correlated hopping term exhibits
the fully opposite effects on both the formation and condensation
of excitons as well as valence transitions. Indeed, Fig.~3a shows 
that the density of zero momentum excitons $n_0$ is now generally
enhanced with increasing $U_{ch}$ (this effect is especially
strong near $E_f=0$) and contrary to the previous case, where
the intermediate phase with $n_d\sim 0.5$ has been stabilized,
it is now suppressed, and due to, the valence transitions become
more steeper. 
\begin{figure}[h!]
\begin{center}
\includegraphics[width=7.0cm]{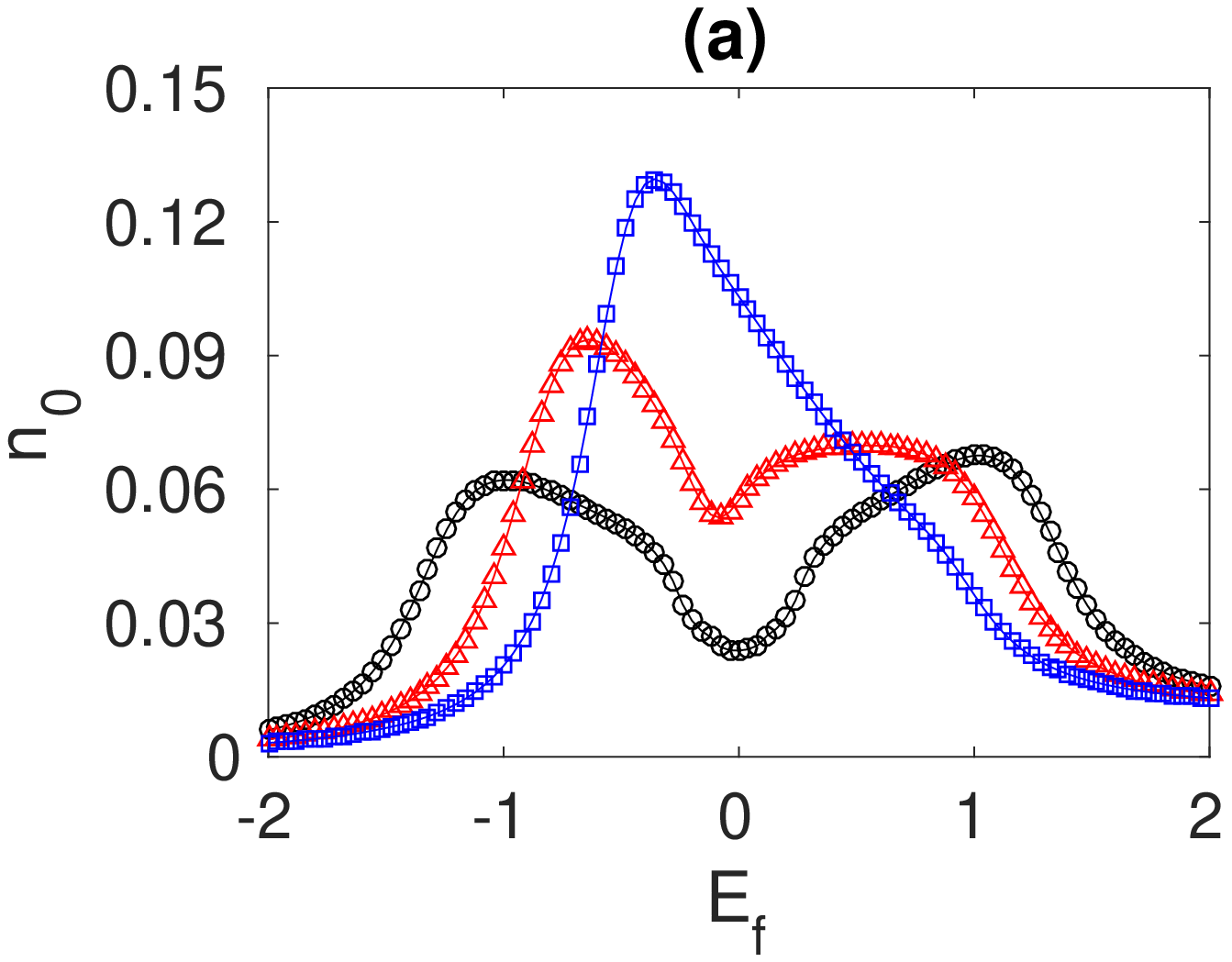}
\includegraphics[width=7.0cm]{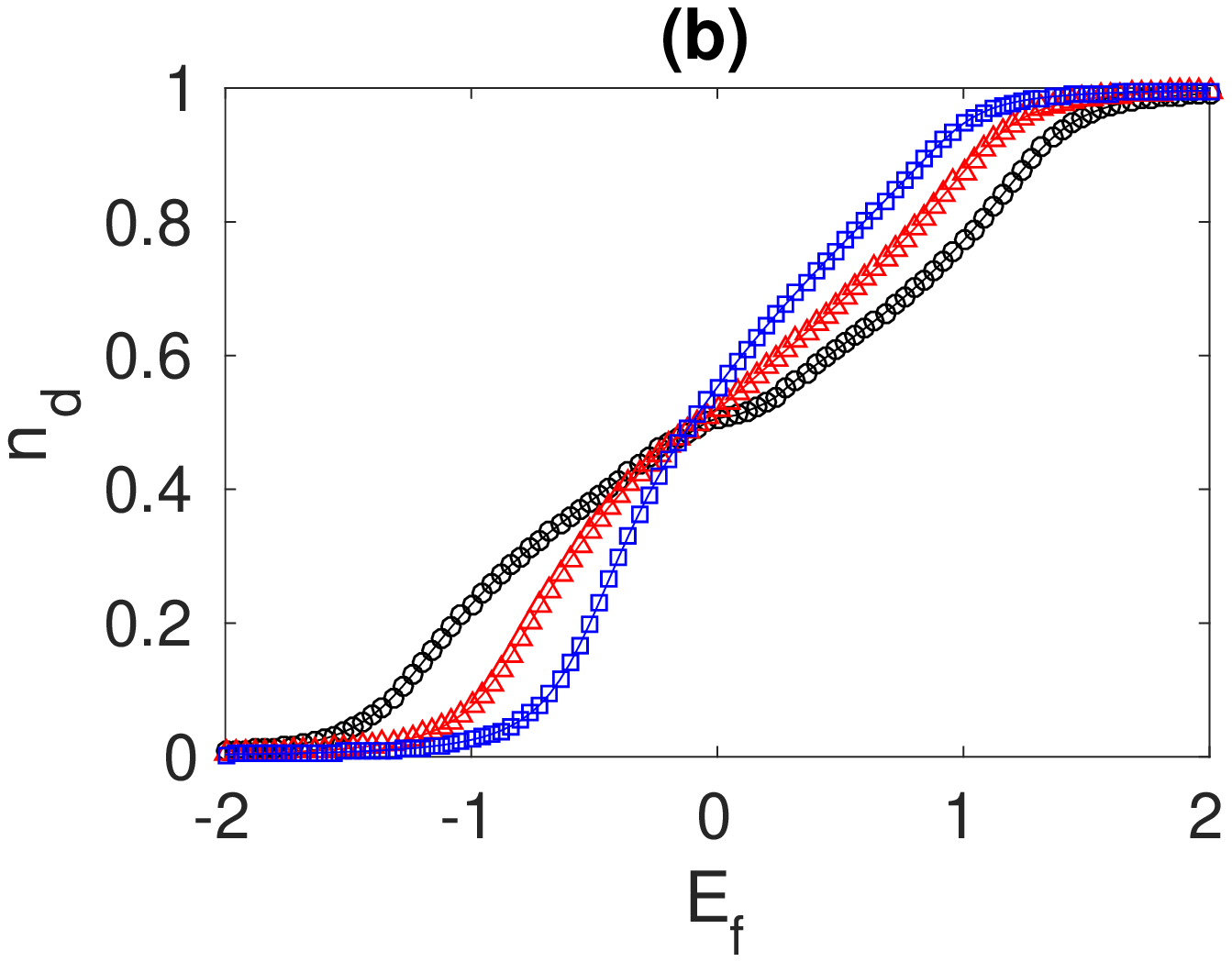}
\includegraphics[width=7.0cm]{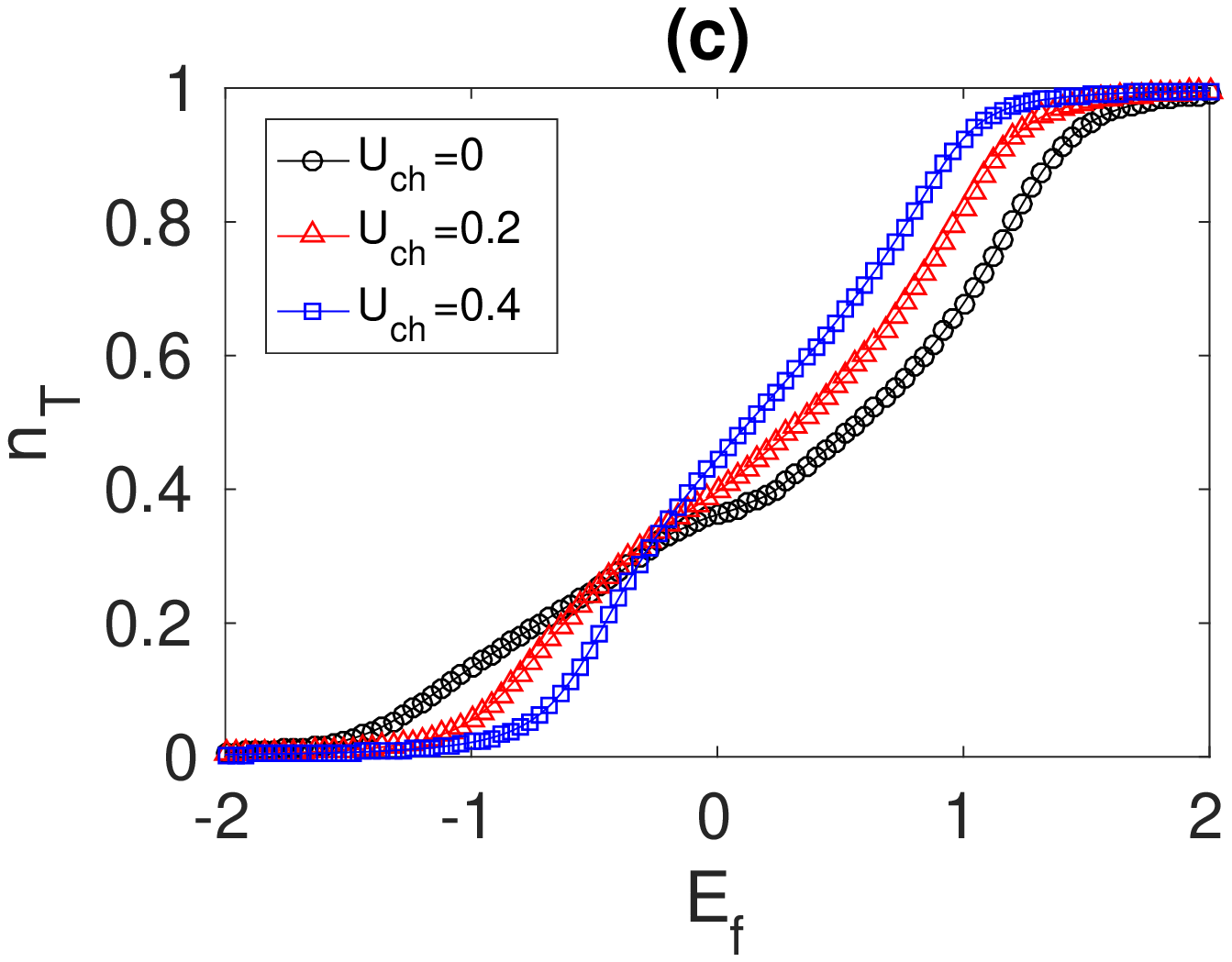}
\includegraphics[width=7.0cm]{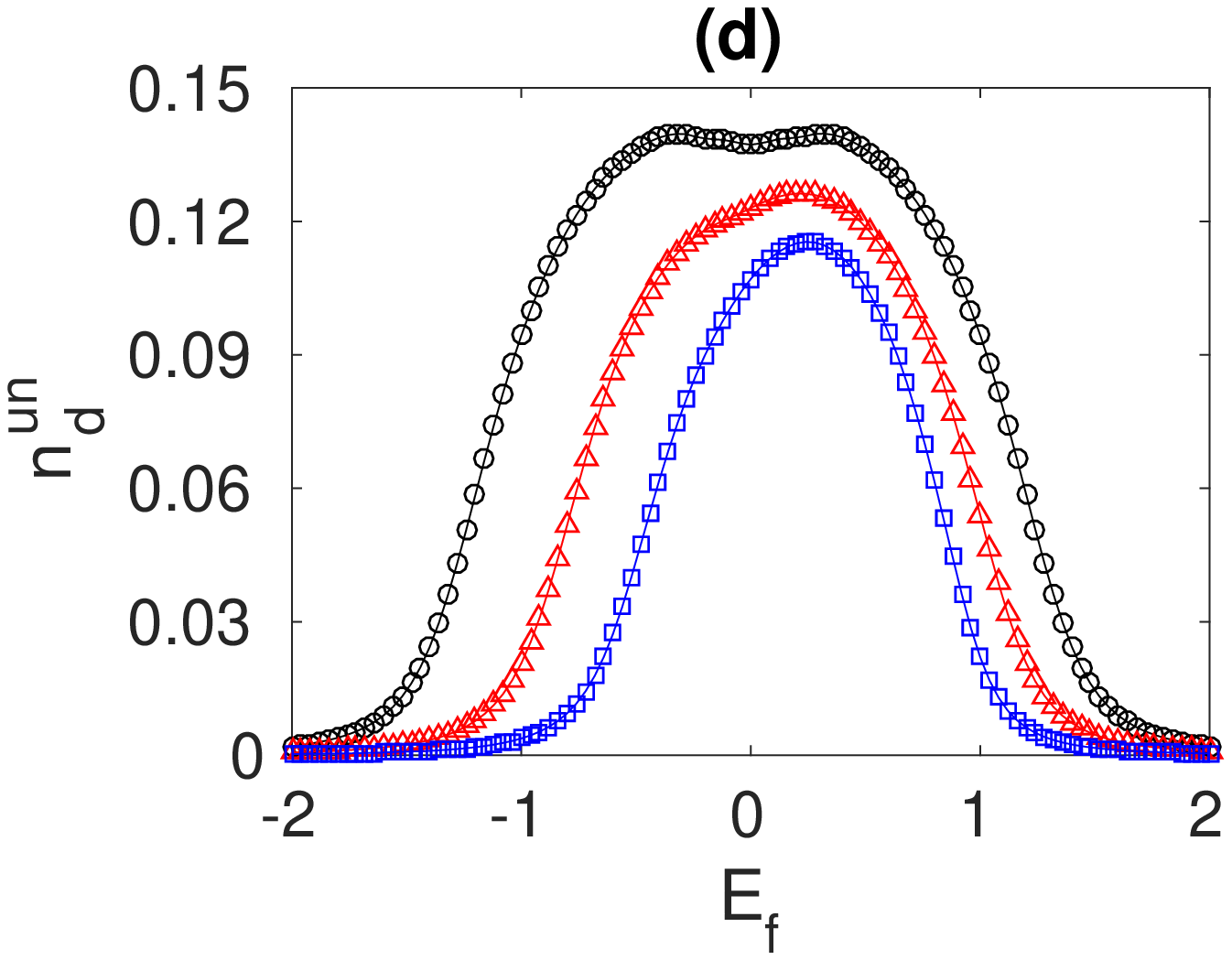}
\end{center}
\caption{\small $n_0, n_d, n_T$ and $n^{un}_d=n_d-n_T$ as functions of $E_f$
calculated for three different values of $U_{ch}$ ($U_{ch}=0, 0.2, 0.4$)
at $U_{nn}=0, U=1, V=0.1, L=100$ and $n_f+n_d=1$.
}
\label{fig3}
\end{figure}
The fully different behaviour (in comparison to the $U_{nn}>0$ case)
exhibits also the total density of unbond $d$ electrons $n^{un}_d$,
which is generally suppressed with increasing $U_{ch}$.

The physically most interesting case corresponds, however, 
to the situation when both ($U_{nn}$ as well as $U_{ch}$)
interactions are switched on simultaneously. The results of our 
DMRG calculations obtained for the case $U_{nn}=U_{ch}$ are 
displayed in Fig.~4. 
\begin{figure}[h!]
\begin{center}
\includegraphics[width=7.0cm]{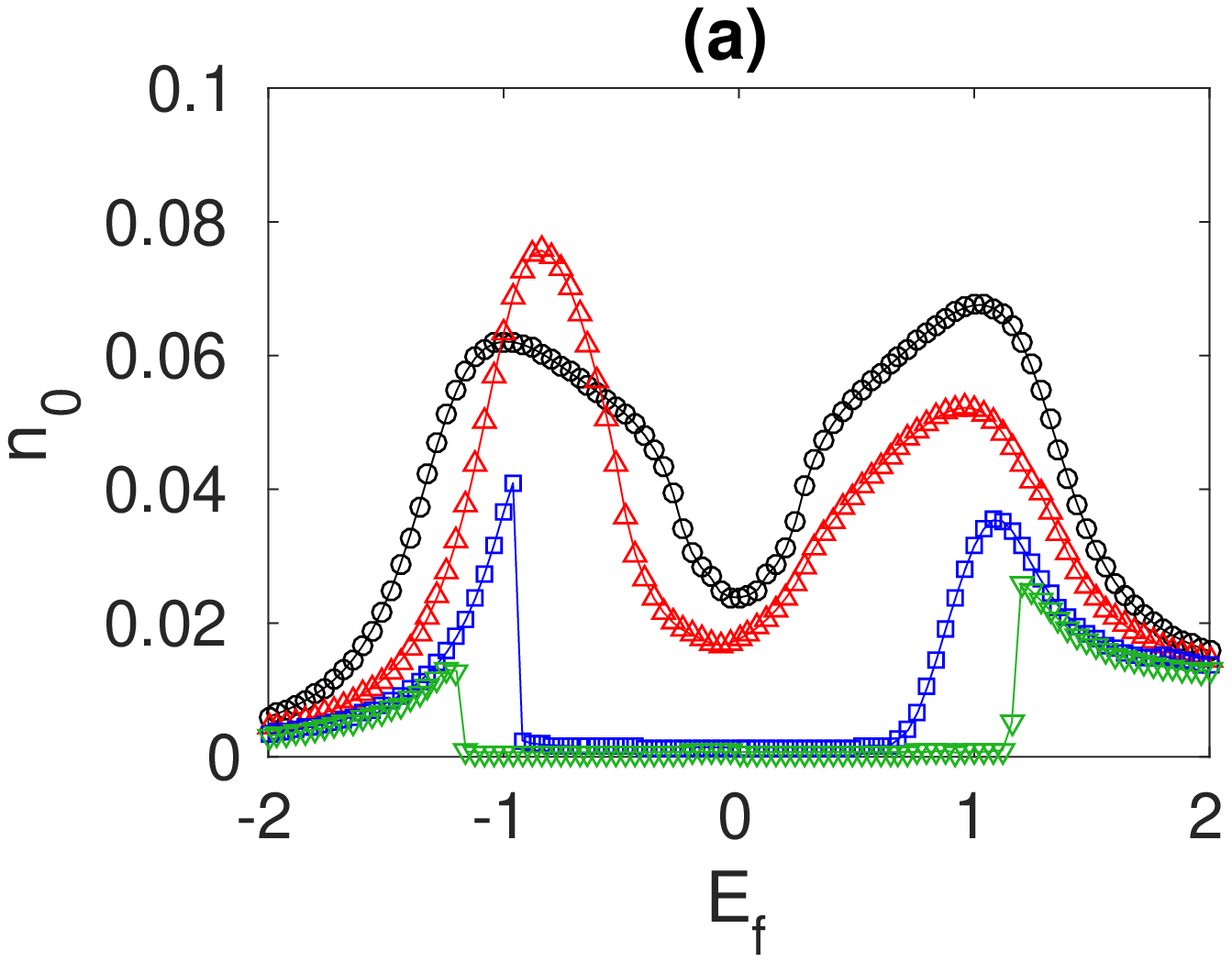}
\includegraphics[width=7.0cm]{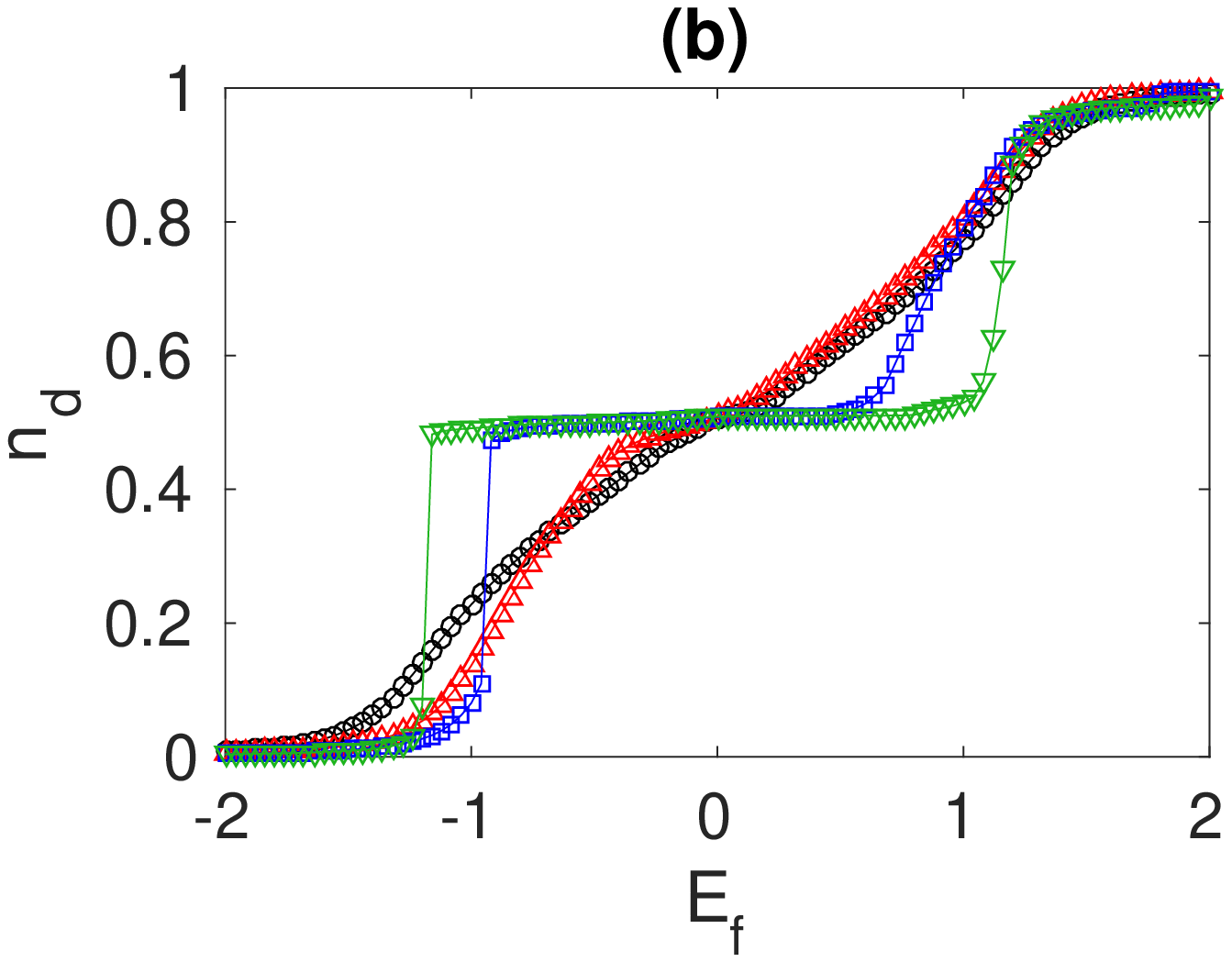}
\includegraphics[width=7.0cm]{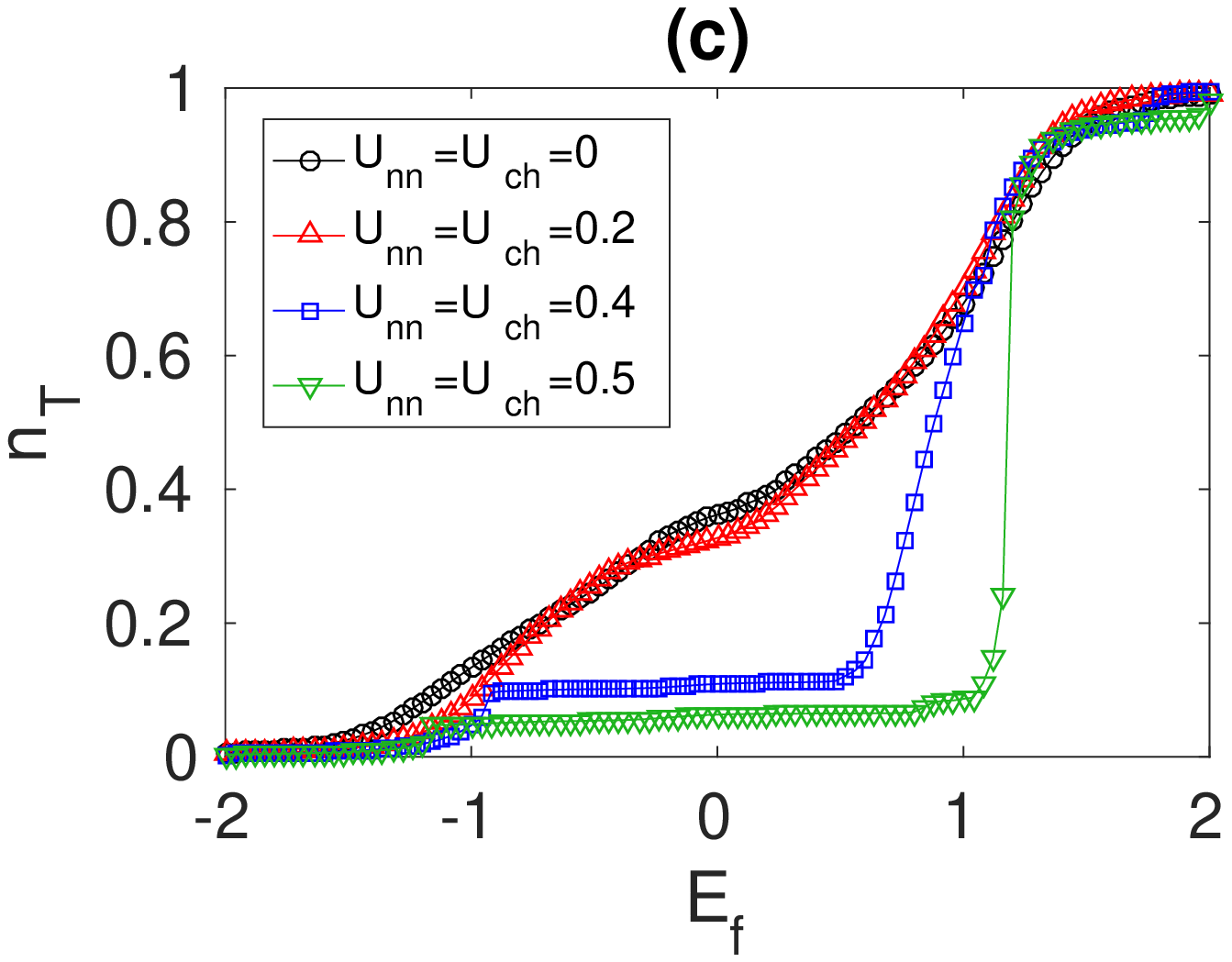}
\includegraphics[width=7.0cm]{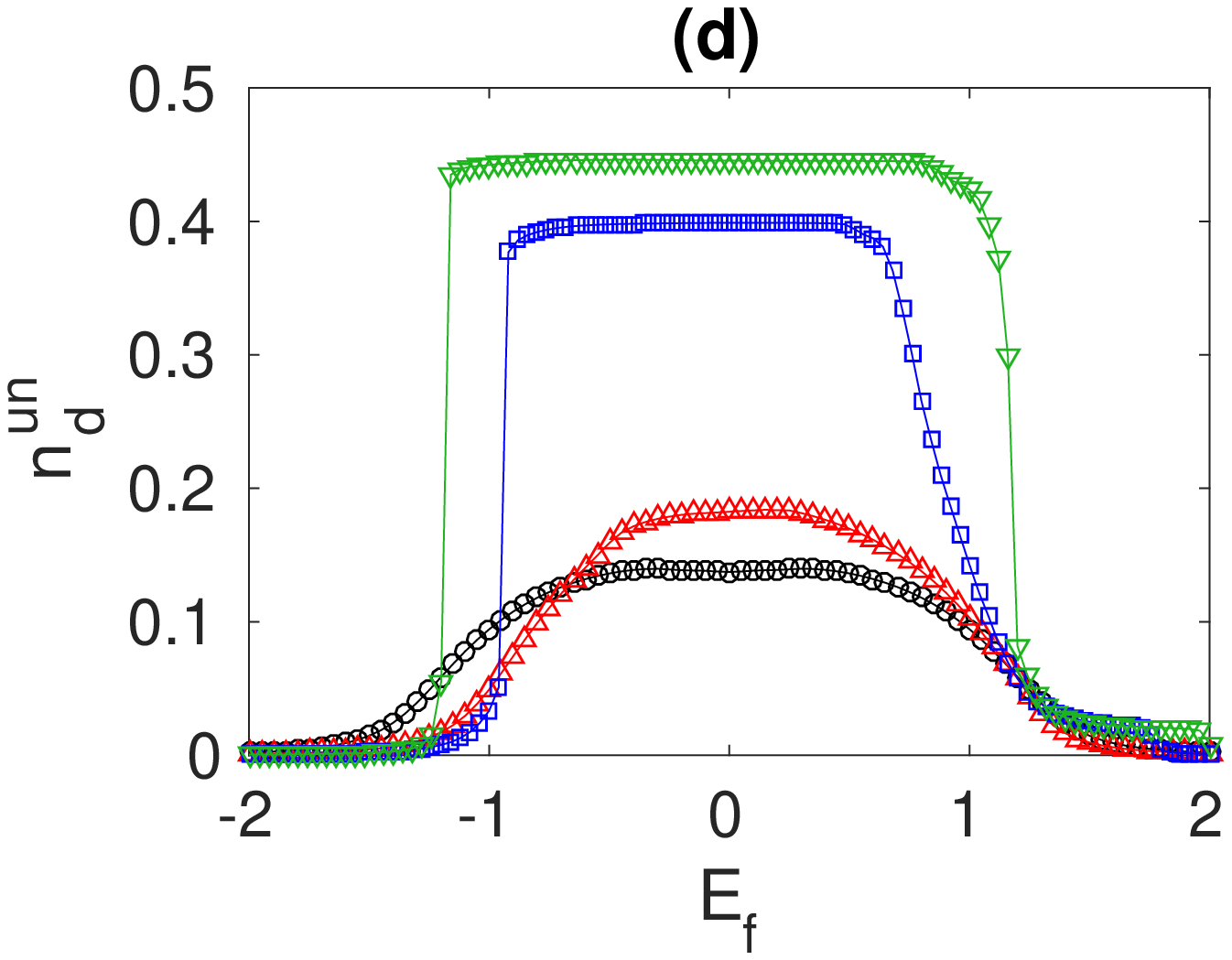}
\end{center}
\caption{\small $n_0, n_d, n_T$ and $n^{un}_d=n_d-n_T$ as functions of $E_f$
calculated for four different values of $U_{ch}$ ($U_{ch}=0, 0.2, 0.4, 0.5$)
at $U_{nn}=U_{ch}, U=1, V=0.1, L=100$ and $n_f+n_d=1$.
}
\label{fig4}
\end{figure}
One can see that there are some   
similarities, but also differences in comparison to the
above discussed cases ($U_{nn} > 0, U_{ch}=0$ and $U_{nn}=0, U_{ch} > 0$).
Let us first discuss the similarities. There are: (i) the strong    
suppression of the zero-momentum condensate in the region of $E_f$,  
where $n_d \sim 0.5$, (ii) the stabilization of the intermediate phase
with $n_d \sim 0.5$ for increasing $U_{nn}=U_{ch}$ and (iii) the 
strong enhancement of the total density of unbond $d$ electrons
$n^{un}_d$ with increasing of $U_{nn}=U_{ch}$. It  should be noted, that 
these similarities with the case $U_{nn}>0, U_{ch}=0$ are quite surprising,
since as discussed above $U_{nn}$ and $U_{ch}$ exhibit fully opposite
effects on the formation of zero momentum condensate, as well as
on the valence transitions. 

More interesting are, however, differences which are produced
by the combined effects of $U_{nn}$ and $U_{ch}$ interaction
in comparison to the cases when $U_{nn}$ and $U_{ch}$ act individually.
There are: the stabilization of zero momentum condensate   
for some values of the $f$-level energy $E_f$ in the 
weak coupling limit $U_{nn}=U_{ch} \sim 0.2$, (ii) the appearance 
of the discontinuous valence transitions for sufficiently large 
values of $U_{nn}=U_{ch} \sim 0.4$, (iii) the discontinuous
disappearance of the density of zero momentum excitons,
as well as the discontinuous changes in the total density 
of excitons $n_T$ and the total density of unbond $d$ 
electrons $n^{un}_d$ at the valence transition points.

The appearance of the discontinuous changes in some ground-state
observables like the density of conduction $d$ (valence $f$) electrons,
the density of zero-momentum condensate, the density of unbond
electrons is very important result from the point of view of 
rare-earth compounds. In some of them, e.g., the mixed valence 
system SmS such discontinuous changes are observed experimentally 
in the density of valence electrons when  the external hydrostatic 
pressure is applied~\cite{SmS}, however, they were not 
satisfactorily described till now. Indeed, as mentioned
above the SmS compound is the mixed valence system, with 
fluctuating valence and thus for its description one has to take
into account the hybridization between the localized $f$ and 
conduction $d$ electron states. However, more reliable methods,
like alloy-analog approximation~\cite{AAA}, renormalization group 
method~\cite{Hanke}, exact diagonalization method~\cite{Fark_zphys} predict
only the continuous valence transitions within the Falicov-Kimball
model extended by the local hybridization. Here we show that considering
the parametrization between the external pressure $p$ and the
$f$-level position $E_f$, the pressure induced discontinuous
valence transitions is possible to generate also in such a system
under very realistic assumption, and namely, that nonlocal interactions
are switched on.  This opens new route to understanding various
ground-state anomalies observed in the rare-earth compounds
within the unified picture. 

In summary, we have used the DMRG calculations to examine the combined 
effects of various nonlocal interactions on valence transitions and the 
formation of excitonic bound states in the generalized Falicov-Kimball model. 
In particular, we have considered the nearest-neighbour Coulomb interaction 
(i) between two $d$ electrons ($U_{dd}$), (ii) between one $d$ and one $f$ 
electron ($U_{df}$), (iii) between two $f$ electrons ($U_{ff}$) 
and (iv) the so-called correlated hopping ($U_{ch}$). 
To reduce the parametric space of the model we have restricted
our study only to the case $U_{dd}=U_{df}=U_{ff}=U_{nn}$ and 
examined the individual and combined effects of $U_{nn}$ and $U_{ch}$ 
on valence transitions and the formation (condensation) of excitonic bound 
states of conduction $d$ and valence $f$ electrons.
Analysing numerically the density of conduction $n_d$ 
(valence $n_f$) electrons and the excitonic momentum distribution
$N(q)$, as functions of $U_{nn}$ and $U_{ch}$ it was found that these
interactions exhibit fully different effects on the valence transitions 
as well as on the process of formation and condensation of excitonic bound
states. While the nonlocal interaction $U_{nn}$ suppresses the formation 
of zero momentum condensate and stabilizes the intermediate
valence phases with $n_d \sim 0.5, n_f \sim 0.5$, the correlated hopping 
significantly enhances the number of excitons in the zero-momentum condensate 
and suppresses the stability region of intermediate valence phases. However, 
the physically most interesting results are observed if both $U_{nn}$ 
and $U_{ch}$ are nonzero, when the combined effects of $U_{nn}$ and 
$U_{ch}$ are able to generate discontinuous changes in $n_f$, $N(q$=$0)$ 
and some other ground-state quantities of the model.

\vspace{0.5cm}
{\small This work was supported by projects ITMS26220120047, VEGA 2-0112-18
and APVV-17-0020. Calculations were performed in the Computing Centre 
of the Slovak Academy of Sciences using the supercomputing infrastructure
acquired in project ITMS 26230120002 and 26210120002 (Slovak infrastructure
for high-performance computing) supported by the Research and Development
Operational Programme funded by the ERDF.}

\end{document}